\documentclass[aps,twocolumn]{revtex4} 
\usepackage{color}
\usepackage{epsfig}
\makeindex

\usepackage{amssymb}   
\usepackage{amscd}   
\usepackage{ulem}
\usepackage{amsmath}

\newcommand{\be}{\begin{eqnarray}}
\newcommand{\ee}{\end{eqnarray}}

\newcommand{\bra}[1]{\left\langle #1 \right|}               
\newcommand{\ket}[1]{\left| #1 \right\rangle}               

\newcommand{\yb}{$^{171}$Yb$^+$}

\newcommand{\beq}{\begin{equation}}
\newcommand{\eeq}{\end{equation}}

\begin{document}
\title{Cooling Techniques for Trapped Ions\label{ch1}}


\author{Daniel M.~Segal}
\affiliation{The Blackett Laboratory, Imperial College London, Prince Consort Road, SW7 2AZ, United Kingdom, \\ d.segal@imperial.ac.uk}
\author{Christof Wunderlich}
\affiliation{Department Physik, Universit\"at Siegen, 57068 Siegen, Germany, \\
wunderlich@physik.uni-siegen.de}


\date{Les Houches Winter School January 2012}

\begin{abstract}
This book chapter gives an introduction to, and an overview of, methods for cooling trapped ions. The main addressees are researchers entering the field. It is not intended as a comprehensive survey and historical account of the extensive literature on this topic. We present the physical ideas behind several cooling schemes, outline their mathematical description, and point to relevant literature useful for a more in-depth study of this topic.
\end{abstract}


\maketitle

\section{Introduction}
\index{laser cooling}
Trapped ions are used in a wide variety of areas of experimental physics from high-precision measurements and metrology through to applications in Quantum Information Processing (QIP) and Quantum Simulation (QS). The key to the success of many of these applications is the ability to cool the trapped ions down to low temperatures (or, for single ions, more correctly, to reduce their kinetic energies to very low values). This article is based on a series of lectures given at a winter school on Trapped Ions held in January 2012 at Les Houches in France. The aim of the lectures was to give an up-to-date account of the cooling techniques that have been employed for trapped ions in the past and to give an overview of new techniques currently being developed. Though other techniques are discussed, the article concentrates on the extremely successful approach of laser cooling (section \ref{sec:laser}).

It is worth first considering how hot ions might typically be when they are first loaded into an ion trap. A typical trap has a depth somewhere in the region of 0.1-100 eV. Ions will remain trapped if they have this sort of kinetic energy when they are near the centre of the trap. Using $E \sim k_BT$ we find that ions with a temperature of $T\sim 1 \times 10^3 $K can easily be trapped.
In comparison to the kinds of traps available for neutral atoms these traps are extraordinarily deep. Furthermore traps can be made to be very small so that oscillation frequencies in ion traps can be very high, easily in the region of tens of MHz.

How hot the ions initially are depends critically on the loading technique. Ions may be either injected into a trap from outside or can be created in-situ from neutral atoms passing through the trap. In the former technique ions are typically loaded in bunches. \index{ion trap loading} A typical trap has a direct
current (DC) voltage applied to some sort of endcaps (which may be hollow to allow the passage of ions). The voltage of one endcap is temporarily brought to a low value while a bunch of ions enters the trap. The endcap voltage is then ramped up quickly so that the ions that bounce off the potential created by the far endcap find themselves trapped. The initial temperature of ions loaded in this way depends on the internal temperature of the initial ion bunch and upon the details of how the electrode pulsing is implemented, but typically initial temperatures are high and  similar to the actual trap depth.  In order to do useful experiments often some form of cooling is required so that the ions become well localized at the centre of the trap rather than filling its entire volume. Of course the density of ions near the centre of the trap is limited by their mutual Coulomb repulsion so that final densities are many orders of magnitude lower than they can be for trapped neutral atoms.

As an example of the injection loading technique, the HITRAP facility at the Gesellschaft f\"ur Schwerionenforschung (GSI) in Darmstadt  \cite{HITRAP} \index{HITRAP} will load a variety of different experimental Penning traps with highly charged ions generated by smashing an energetic beam of heavy ions into a stripping target. Exotic ions up to and including hydrogen-like uranium (a uranium nucleus with only a single orbiting electron remaining) can be created in this way. After many stages of slowing and pre-cooling, bunches of highly charged ions at a temperature of around 4K will be directed into an experimental trap. One such trap is operated by the SPECTRAP collaboration whose aim is to use cold trapped highly charged ions to make optical spectroscopic measurements of hyperfine transitions that, in ordinary ions, would be in the radio or microwave region of the spectrum. These experiments will allow very sensitive tests of QED in situations where electric and magnetic fields are huge so that perturbative approaches to QED are stretched to their limits. Despite the initial internal temperature of the ion bunch being quite low it is expected that imperfect control of the transport of these ions into the final trap and `shutting the door' on them will lead to relatively high initial temperatures in the trap in the region of an eV or more. A variety of in-trap cooling techniques will therefore be brought to bear once the bunch of highly charged ions has been captured.

\section{Non-Laser Cooling Techniques}
\subsection{Electron Cooling}

\begin{figure}
\centerline{\psfig{file=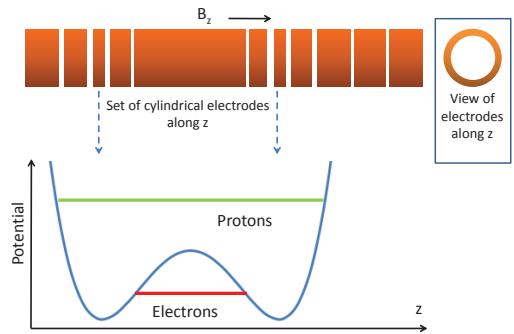,width=7cm}}
\caption{Multiple electrodes along the trap axis allow complicated potentials to be created. A W-shaped potential can hold a small cloud of electrons in the middle of a larger cloud of protons. Recombination (to form neutral H which would then be lost from the trap) is surprisingly rare so that elastic collisions dominate leading to thermalisation between the electrons and the protons. The electrons in their cyclotron orbits radiate rapidly removing kinetic energy from the system leading to `sympathetic' cooling of the protons.  Note - for the antihydrogen work recombination is the goal so strategies for enhancing this process are necessary (adapted from reference~\cite{hall1996}).}
\label{electroncooling}
\end{figure}

The pre-cooling of ions in the HITRAP facility will be accomplished using electron cooling\index{electron cooling}. This is a technique often used in so-called `nested' Penning traps. A Penning trap can be comprised of a series of hollow cylindrical electrodes. By making a structure with a large number of such electrodes along a line, an array of traps for positively and negatively charged particles can be formed. By generating a W-shaped potential as shown in figure \ref{electroncooling} electrons and positive ions can be trapped in the same region of space where they can interact (the electrons are confined to the inverted well owing to their negative charge). An interesting example of this sort of cooling is in the field of Anti-Hydrogen production. In these experiments negatively charged anti-protons are cooled through collisions with positrons (see reference\cite{Madsen2013} and the web pages of the ALPHA, ASACUSA and ATRAP collaborations at CERN).

As a precursor to these experiments, Hall {\em et al.}\cite{hall1996} performed a pilot experiment in order to develop the technique, in which protons were cooled through collisions with electrons. In their experiment the cylindrical electrodes could have voltages up to 150 V, their magnetic field was B=6 T and the trap was held in a cryostat at T=4.2 K. A 40 nA, 1 kV electron beam from a field emission point was fired along the B-field towards a metal plate at the end of the trap. Hydrogen ejected from the plate was then ionised by the electrons.  The protons generated were initially trapped in a shallow harmonic well (to the right of the W-shaped potential shown in figure~\ref{electroncooling}) formed by applying appropriate potentials to the trap electrodes. The protons could then be loaded into the W shaped potential well with a well known initial energy. The protons could then be ejected from the trap and detected. By measuring the profile of their time-of-flight a temperature for the protons could be inferred. This temperature was found to be commensurate with the energy they had when they were injected into the trap. It was also possible to use the electron beam to pre-load some electrons into their trap. The electrons thermalise with the 4.2K cryogenic background in about 0.1s by emitting synchrotron radiation. Elastic collisions with the protons cool the protons and when they were subsequently ejected they were found to emerge at a significantly reduced temperature.

\subsection{Resistive Cooling}

\index{resistive cooling} The idea here is to couple a trapped ion to an external circuit in such a way that the energy is dissipated in the circuit, cooling the ion. Following the approach outlined by Holzscheiter \cite{holzscheiter1988} we imagine a charged particle  (mass $m$, charge $q$) oscillating between a pair of infinite parallel plates separated by $2z_0$ connected by a resistor $R$. The oscillating particle induces an image charge given by \cite{holzscheiter1988}

\begin{equation}
q^\prime=(z_0\pm z)q/2z_0
\end{equation}

This induces a  current in the resistor

\begin{equation}
i=qv_z/2z_0
\end{equation}

As a result energy is dissipated at a rate

\begin{equation}
dE/dt= -i^2R = q^2RE/4mz_0^2
\end{equation}

This in turn leads to exponential cooling of the charged particle with a time constant given by \cite{holzscheiter1988}

\begin{equation}
t_{\rm nat}= 4mz_0^2/q^2R.
\end{equation}

If the resistor is cooled for instance by placing it in a cryogenic environment then eventually the charged particle will come into thermal equilibrium with the resistor. Thus, this technique can in principle result in temperatures in the region of 4K (liquid He), and, since initial ion temperatures can by as high as the trap depth, resistive cooling is often well worth implementing. In practice the technique is usually operated in a resonant fashion whereby the circuit connected to the ion is not simply resistive but has some reactance (in this case the external circuit is known as a `tank circuit'). While this in principle makes the technique more effective it adds a complication for clouds of ions where many resonant frequencies may be apparent due to the many modes of oscillation of the ion cloud. Resistive cooling also requires cryogenic operation which has a clear experimental overhead. As a result, unless cryogenic operation is required for some other reason, it is worth considering other cooling methods.

\subsection{Buffer Gas Cooling}

\index{buffer gas cooling} Again, bearing in mind the very high temperatures that newly loaded ions may have in traps, even cooling the ions down to room temperature could be very beneficial for some experiments. In radiofrequency (RF) traps this can be accomplished by allowing a small amount of neutral `buffer gas' into the trap along with the trapped ions. The buffer gas particles, which are in thermal equilibrium with the trap container through frequent collisions with the walls, collide with the ion, removing kinetic energy and transporting it to the walls of the chamber as heat. To avoid chemical reactions occurring between the neutral particles and the ion, the buffer gas is chosen to be inert and typically the lightest noble gas available, helium, is used. A pressure of He in the region of $10^{-6}$ mbar suffices. This technique is used extensively in Linear Ion Trap Mass Spectrometry to cool samples and increase sensitivity.

Buffer gas cooling in a Penning trap is possible but requires an extra technique to be brought to bear. In a Penning trap a static electric quadrupole potential is used to confine the ions in the axial direction. This static potential actually repels ions away from the centre of the trap towards the walls. The radial confinement in this trap comes from the imposition of a strong axial magnetic field. As a result the ions move around the top of a radial electrostatic potential hill in circular `magnetron' orbits. On a smaller scale the ions perform cyclotron orbits at a frequency that is reduced by the presence of the electric potential -- the `modified cyclotron frequency'.  If energy is taken away from the ions by collisions with the buffer gas then they will tend to move down the electric potential and thus towards the walls of the trap.

The magnetron motion is therefore unusual insofar as it is unstable. The modified cyclotron motion is `normal' in the sense that removing energy from it results in smaller cyclotron loops.  Buffer gas cooling can be made to work in the Penning trap by a trick that couples the `normal' modified cyclotron motion to the unstable magnetron motion. In this way energy is transferred periodically between the two motions. While the ion's motion is predominately cyclotron-like the the ion is cooled strongly by the buffer gas collisions. For the right parameters this strong cooling can be made to dominate over the tendency of the ions to move outwards in the trap when the ion is moving in a more purely magnetron-like orbit \cite{buffergas}.

\section{Laser Cooling}
\label{sec:laser}
\index{laser cooling} The idea of using lasers to cool samples of atomic particles was independently suggested for ions \cite{Wineland1975} and for neutral atoms \cite{Haensch1975} in 1975. The first experimental demonstrations of laser cooling of trapped atomic ions came a few years later in 1978 \cite{Neuhauser1978,Wineland1978} followed soon by a thorough theoretical analysis specifically for trapped ions \cite{Wineland1979}. Laser cooling for a single ion was demonstrated in 1980 \cite{Neuhauser1980}. Since these early days laser cooling has become an ubiquitous technique used extensively to produce extremely cold samples of both ions and neutral atoms. Some of the main ideas have been covered extensively elsewhere in the literature  (e.g., \cite{adamsriis,winelanditano,Stenholm1986,Itano1995,Eschner2003}), and we will limit ourselves to giving a basic description of the process here. However, there have been a number of recent developments specific to ion traps and we will focus more closely on these.

When an atom (or ion) absorbs a photon from a laser beam the momentum of the photon $\hbar k$  is imparted to the atom ($k$ is the wavenumber). For visible photons  $\hbar k$ is small, in the region of $ 1\times 10^{-27}$kgm/s. Though this is indeed small, it is not entirely negligible compared to the relatively small momentum of a room temperature atom. If the laser beam is directed against the direction of motion of the atom this momentum kick slows the atom down. Laser cooling harnesses this interaction to bring gaseous samples of atoms or ions down to very low temperatures. However, there are some difficulties that need to be overcome. Once the atom has absorbed a photon it is in an excited state. After a short time (typically a few ns) the atom re-emits the photon spontaneously, receiving another momentum kick in the process. The atom is then ready to go around the cycle of absorption and emission again. At first sight it would seem that nothing has been gained, but closer consideration shows that the momentum kicks from the absorption events are additive, since all the photons involved come from the laser beam, whilst the spontaneously emitted photons have a nearly isotropic distribution, so that the recoil momentum kicks tend to cancel each other out over many cycles of the process.  The fact that atoms can absorb and emit photons so rapidly means that scattering can provide an enormous deceleration to a moving atom -- in the region of $10^5 g$ ($g$ is the earth's gravitational acceleration). However, in order to realise the potential of this force the Doppler effect must be taken into account.

\index{Doppler cooling} Imagine a beam of atoms moving towards an oncoming laser beam. The problem is that atomic transitions are relatively narrow in frequency. In order for laser cooling to work the laser must be deliberately detuned to the red side of the natural transition frequency of the atom. In this way the moving atom is shifted into resonance with the laser and can scatter a photon. The next problem is that as an atom slows down, the Doppler effect takes it out of tune with the incoming radiation, potentially switching the process off after a few scattering events. It is easy to show that of the order of $10^4$ scattering events are required to bring a room temperature atom to rest. There has been an enormous amount of activity in the area of laser cooling of neutral atoms and a variety of techniques have been developed that overcome the problems described above allowing large numbers of scattering events to take place and eventually leading to ultra-cold samples of neutral atoms which can be stored in a variety of neutral atom traps. Most of these techniques are not relevant to trapped ions. However it is important to understand that this type of cooling -- called Doppler cooling -- whilst being extremely effective, does have its limitations. In particular the description above assumes that atoms only have two energy levels. This is a gross simplification and in real atoms multiple laser frequencies are often required in order to pump atoms out of metastable energy levels where they could otherwise become stranded for long periods of time.

\index{Doppler cooling limit} Although very low temperatures can be achieved using Doppler cooling there is a limit -- the Doppler limit -- and this limit {\em is} relevant to trapped ions. The argument above is rather too simplistic to capture the physical process that leads to this limit. Firstly, even a stationary atom does not absorb light at a single well-defined frequency -- the atomic transition has a `lineshape' due to natural broadening that is Lorentzian. Secondly, the atoms or ions one wishes to cool do not typically all move in a set direction but instead move about randomly undergoing collisions with each other. At any given time some atoms in the gas will be moving towards the red-detuned laser and will thus be more likely to scatter a photon, however even atoms that are moving away from the laser beam will have some probability of absorbing a photon in the wing of the absorption line. For neutral atoms a strategy often employed is to use pairs of counter-propagating laser beams so that whatever direction the atom may be moving in, it will find itself Doppler shifted into resonance with a laser beam that tends to slow it down. By considering the scattering of photons from atoms moving in one dimension in counter-propagating red detuned laser beams at this level of complexity it can be shown that the radiation pressure or `scattering force' $F$ is given by

\begin{equation}
F \sim \hbar k^2 v
\label{eq:force}
\end{equation}
where $\sim$ indicates that the r.h.s of equation~\ref{eq:force} is actually preceded by a numerical factor. If the laser power is low so that the transition is not overly saturated and the detuning is of the order of the natural linewidth of the transition $\gamma$ then this numerical factor will be of order unity.  This is a velocity-dependent force and can therefore be viewed as a form of friction. This friction removes energy from the atom at the rate

\begin{equation}
-F.v \sim \hbar k^2v^2
\end{equation}

In order to understand the Doppler limit it is necessary to look back at the argument above concerning the near-isotropic emission events causing momentum kicks that tend to cancel each other out over many absorption-emission cycles. Although this is broadly true, the recoil momentum kicks during spontaneous emission cannot be ignored completely. If we imagine an initially stationary atom undergoing this process in isolation we see it making a random walk in momentum space. If the atom can scatter photons at a rate $R$ photons per second, then its average momentum will grow as

\begin{equation}
\Delta p=\sqrt n \hbar k=\sqrt{Rt}\hbar k
\end{equation}

Again, for well chosen values of the detuning and the saturation parameter the scattering rate can be of order $\gamma$. We will henceforth replace $R$ with $\gamma$ but bear in mind that the equations that follow are only approximate. Over a time $t$, $n$ photons will be  scattered and the average increase in kinetic energy due to the random walk is

\begin{equation}
{\Delta p^2 \over 2m} = {\hbar^2 k^2 \gamma t \over 2m}
\end{equation}

This means that the random spontaneous scattering of photons leads to a heating rate given by $\hbar^2 k^2\gamma/2m$. The limit to the achievable temperature can be found by equating this heating rate to the cooling rate and finding out what temperature results from this equilibrium:

\begin{equation}
{ \hbar^2 k^2 \gamma \over 2m} = \hbar k^2 v^2 \  ,
\end{equation}

so that

\begin{equation}
v^2={\hbar \gamma \over 2m} \  .
\end{equation}

The average kinetic energy in one dimension realised in this way can be related to temperature through

\begin{equation}
\langle \frac{mv^2}{2}\rangle =\frac{\hbar \gamma}{4}=\frac{1}{2}k_BT
\end{equation}

giving the famous \index{Doppler cooling limit} Doppler limit \footnote{This simple approach captures the basic physics and produces the Doppler limit formula. That it includes the well known factor of two in the denominator is fortuitous since it leaves out a number of important subtleties. It is based on a more rigorous treatment outlined in a series of masters level lectures on laser cooling of neutral atoms given by Prof. E.A. Hinds. For a more complete treatment which is specifically aimed at the special case of trapped ions see reference \cite{Wineland1979}} 

\begin{equation}
\frac{\hbar \gamma}{2} = k_BT \  .
\end{equation}

Amongst the first neutral atoms to be cooled in this way were Rb and Na for which the Doppler Limits are 144\,$\mu$K and 240\,$\mu$K respectively (using the D line for laser cooling), astonishingly low temperatures by any reckoning. The ingenuity of workers in the field of laser cooling has led to a number of strategies for circumventing the Doppler Limit and reaching even lower temperatures. Most of these techniques are of limited relevance to trapped ions for which alternative `sub-Doppler' cooling techniques have been developed. These techniques will form the focus of the rest of this article. To understand these techniques it is important to ask the question: what is fundamentally the lowest temperature that a collection of trapped ions can reach? It is important to remember that the ions are held in a harmonic potential well. At some level the quantisation of the motion of the ion in this potential well must become manifest. Indeed the lowest temperature that the ions can reach is set by the zero-point or motional ground state energy of an ion held in this potential well. Since ion traps can be made so tightly confining, the ultimate temperature that trapped ions can reach is higher than the lowest temperatures achievable for trapped neutral atoms where the traps are in effect comparatively weakly confining. Nonetheless, as we will see, it is possible to cool trapped ions to the extent that they spend the vast majority of their time in the motional ground state and in this state the ions are as cold as they physically can be.

Trapping neutral atoms is often  achieved using a so-called Magneto-Optical Trap (MOT). In this device a number of laser beams derived from the same laser intersect over a small volume in a vacuum chamber where there is a spatially varying magnetic field. A combination of the laser detuning and the Zeeman effect ensure that an atom  feels a scattering force directed towards the centre of the volume.
This combination of a static magnetic field landscape and the  light field lead to a velocity and position dependent force, and thus forms a `trap' for atoms.

By contrast ions, being charged, can be trapped even without the imposition of a light field, and even when they are very hot. Laser cooling can then be applied as a separate consideration. This means that a trapped ion can in principle be laser cooled using a single laser beam. Indeed some species of ion (notably Be$^+$ and Mg$^+$) really do require only a single laser frequency and can be cooled with a single beam. These ions are unusual in that they do not have metastable levels below the first electronically excited state used for laser cooling.  When such metastable levels are present the ion can be lost to them for long periods of time. The requirement for three dimensional cooling is fulfilled by ensuring that the single laser beam is directed into the ion trap making an appreciable angle to all the principle axes of the trap.

It is important to realise that a single trapped ion does not move randomly in an ion trap -- it oscillates harmonically at a set of well defined  motional frequencies. For Doppler cooling the normal situation is that these motional frequencies are much lower than the photon scattering rate. The strategy for laser cooling is therefore straightforward. A laser beam is sent into the trap and detuned to the red of the natural frequency of the ion. Most of the time the ion does not interact with the laser, but the ion's oscillations in the trap periodically bring it into resonance with the laser due to the Doppler effect. During brief intervals within each oscillation period the ion scatters some photons. Because of the judicious red detuning, this only occurs when the ion happens to be moving towards the laser beam. The ion thus slows down and its oscillatory amplitude is reduced. Even if only a few photons are scattered per oscillation cycle the ion gradually loses momentum in the trap and its motion is reduced. For a cloud of ions this constitutes cooling. Collisions between the ions complicate the overall motion but don't change the essential outcome \cite{Stenholm1986}.

Despite the fact that, for ions, trapping and laser cooling are really separable techniques, and despite the significant differences in the details of how laser cooling is applied for ions, the physics behind the Doppler limit is unchanged so that this limit also applies to trapped ions. The rest of this article concerns a variety of techniques that have been developed, or are currently under development that allow cooling well below the Doppler limit to the quantum mechanical ground state of the motion in the trap potential.

\begin{figure}
\centerline{\psfig{file=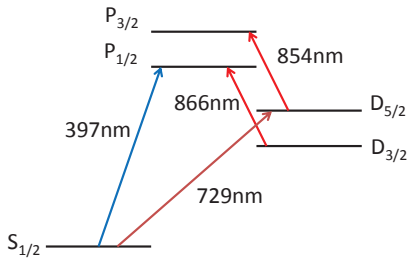,width=7cm}}
\caption{Ca$^+$ level scheme.}
\label{Ca+LevelScheme}
\end{figure}

So far we have considered an idealised atom or ion with only two energy levels, the ground state and a single excited state. Most trapped ions are significantly more complicated than this. The singly charged positive ions of the alkaline earths are the most commonly used ions for laser cooling. This is because the neutral atoms have two electrons outside full shells so that with one electron removed their spectra become particularly simple due to them having a single electron in the outer shell. They thus have spectra that resemble those of the neutral alkalis (which are the most popular neutral atom species for laser cooling). \index{laser cooling of Ca$^+$} $^{40}$Ca$^+$ is one of the most commonly used ion-species, largely because the laser frequencies required are all obtainable using solid state lasers. A simplified energy level diagram for Ca$^+$ is given in figure \ref{Ca+LevelScheme}. Laser cooling is performed on the dipole allowed  $S_{1/2}$ -- $P_{1/2}$  transition at 397 nm. The ion can spontaneously emit to the $D_{3/2}$ level which is metastable, having a lifetime of $\sim$ 1s. After scattering about 16 photons the ion will end up in the $D_{3/2}$ level, switching off the laser cooling for around a second. The result is that the ion would appear dark most of the time and cooling would be very ineffective. To avoid this a `repumper' laser resonant with the $D_{3/2}$ -- $P_{1/2}$ transition at 866 nm is shone into the trap so that the ion is rapidly put back into the $S$ to $P$ cooling cycle as soon as it drops into the $D_{3/2}$ level. The 397nm laser is red-detuned by around a half a natural linewidth to provide laser cooling while the 866 nm laser is tuned to the centre of the  $D_{3/2}$ -- $P_{1/2}$ transition.

\index{sympathetic cooling} One of the limitations of laser cooling is the large experimental overhead involved in setting up to cool a new species of ion.  This typically requires a suite of lasers, all of which need to be controlled carefully. Fortunately, the long range of the Coulomb interaction between ions furnishes us with a very effective means of cooling a very wide range of ions that cannot be directly cooled with lasers. The approach is to co-trap laser cooled ions with other ions which may be of interest for other purposes. These other ions undergo long-range Coulomb collisions with the laser cooled ions and are thus `sympathetically' cooled in the process. These other ions may be different atomic ions or even molecular ions. Provided the mass difference between the laser cooled species and the species to be sympathetically cooled is not too large, the cooling can be very efficient. If the mass difference is very large then the efficiency of the process is lost because the different species of ions settle into positions in the trap that are rather well separated, so that the interactions between the directly cooled ions and the sympathetically cooled ones become weak. In an extreme example of this process a small cloud of around 200 organic molecules of mass 410 Da were sympathetically cooled to a 115mK by a cloud of Ba$^+$ ions \cite{Schiller2006}.

\index{Doppler cooling} For a typical miniature radiofrequency ion trap, Doppler cooling leads to low enough temperatures that the ion's motion in the trap is manifestly quantum mechanical. In general this motion is a mixed state of the harmonic oscillator states of the ion in the trap potential. These states are characterised by a principal quantum number $n$ and values of $\overline n \sim 10$ are quite normal, with lower values down to $\sim$ 1 having been achieved for very small, steep traps\cite{winelandquarterwaveresonator}. So although Doppler cooling is extremely effective, it is not normally capable of cooling ions into the quantum mechanical ground state for which $n=0$. This is because of a `catch 22' situation: the Doppler limit is higher for transitions with high scattering rates. So one strategy might be to simply laser cool on a narrower transition with a lower scattering rate. However for such a transition the cooling rate is also very low. If there are other heating effects present, then achieving very low temperatures will be difficult, if the cooling rate is too low. In fact at this level there are indeed a number of heating effects in ion traps (some of which are not well understood) which are very difficult to control. For this reason a different strategy is needed to go beyond the Doppler limit. However it is important to realise that even though strategies do exist to achieve this, Doppler cooling is always used to pre-cool ions near to the Doppler limit before engaging these other  approaches. Since the motion of a trapped ion at low temperatures is manifestly quantum mechanical, the starting point for understanding sub-Doppler cooling in ion traps must be to write down the Hamiltonian for the system of a trapped ion interacting with a laser field.

\index{laser cooling of Ca$^+$} For reasons hinted at above, a narrow transition will be required in order to perform sub-Doppler cooling of the trapped ion. Fortunately, a very narrow transition is available, for example, in Ca$^+$ which is the $S_{1/2}$ -- $D_{5/2}$ transition shown in figure~\ref{Ca+LevelScheme}. This transition is dipole forbidden but allowed as an electric quadrupole transition. The $D_{5/2}$ state decays to the ground state with a natural lifetime of around 1s. It can be driven using a laser at 729 nm. A few microwatts at 397 nm is all that is required to saturate the normal Doppler cooling transition. In order to achieve excitation within a few ms on the 729 nm transition a great deal more laser power is required to oversaturate this weak transition. Typically a few milliwatts are required, however, this is easily within the reach of diode laser based systems and Ti:Sapphire laser systems. The basic strategy for ground state cooling is to pre-cool the ion to the Doppler limit using the 397\, nm and 866\, nm lasers and then to switch to a different cooling scheme involving the narrow transition at 729\, nm to remove the last few motional quanta from the system leaving it in the ground state of its motion in the trap.

\subsection{Ion-Laser Interaction Hamiltonian}
\index{ion-laser interaction Hamiltonian}
Although the ion motion must be treated quantum mechanically, for most purposes we will find that a classical treatment of the radiation field will suffice. We will assume the ion is a two-level atom (TLA) with a ground state $\vert g\rangle$ and an excited state $\vert e \rangle$. It should be borne in mind that the ion will be cooled to the Doppler limit before beginning the sub-Doppler cooling process we are about to discuss. This process involves the narrow  $S_{1/2}$ -- $D_{5/2}$ transition so that we may identify $\vert g\rangle$ with $S_{1/2}$ and $\vert e\rangle$ with $D_{5/2}$. In what follows, we consider coherent deterministic (de-)excitation of the ion by the laser field only. For laser cooling, as discussed here, in addition a dissipative process is required in the form of spontaneous decay. This ensures that entropy is removed and the ion is cooled. An adequate  mathematical treatment that includes spontaneous emission would require solving a master equation (e.g., \cite{Stenholm1986}). However, the treatment given below is sufficient to gain physical insight into different mechanisms that lead to cooling in the remainder of this chapter.

More detailed developments of the material presented in this section are available in a number of places (e.g., \cite{Loudon,Wineland1998,Major}).

We assume the atomic wavefunctions are known so we can write the time independent Schr\" odinger equation for an isolated atom (ion) as
\beq
H_A\vert i \rangle = \hbar \omega_i \vert i \rangle\,\,,\,\, i=g,e
\label{eq:atomic_hamiltonian1}
\eeq

\noindent
Using the closure theorem and orthonormality of $\vert i\rangle$ allows us to write the atomic part of the Hamiltonian as
\beq
H_A = \hbar \omega_g \vert g\rangle \langle g \vert + \hbar \omega_e \vert e\rangle \langle e \vert
\eeq
\noindent
We define the zero of energy as being midway between $\hbar\omega_g$ and $\hbar \omega_e$ such that $\omega_e=\omega_a/2$ and $\omega_g=-\omega_a/2$
giving
\beq
H_A=\frac{\hbar\omega_a}{2}\{ \vert e\rangle\langle e\vert - \vert g \rangle\langle g \vert \} = \frac{\hbar\omega_a}{2}\sigma_z
\eeq
\noindent
Where we have implicitly defined $\sigma_z$ which is one of the Pauli matrices.
The full Hamiltonian we are interested in must take into account both the ion's motion in the trap and its interaction with an applied laser field linking $\vert g\rangle$ and $\vert e \rangle$. We therefore write
\beq
H=H_A + H_m + H^{(i)}
\label{fullham}\eeq
\noindent
Where $H_m$ relates to the motion of ion of mass $m$ in a 1-D harmonic trap and $H^{(i)}$ relates to the ion's interaction with the laser field. $H_m$ can be written
\beq
H_m=\frac{p^2}{2m} + \frac{1}{2} m \nu^2 z^2 = \hbar \nu (a^\dagger a + 1/2)
\eeq
where $a$ and $a^\dagger$ are related to $z$ and $p$ through the usual relationships for a harmonic oscillator
\beq
z  =\sqrt{\frac{\hbar}{2m\nu}}(a + a^\dagger),  \quad\quad p  = \sqrt { \frac{\hbar m \nu}{2}} (a^\dagger -a)
\label{eq:xandp}
\eeq
and $\nu$ is the angular frequency of the harmonic oscillation. We have in mind a single ion or a small chain of ions in a linear trap for which the harmonic motion along the trap axis $z$ has a lower motional frequency (weaker binding) than the other two axes of the trap. We will only consider the motion in this one dimension. The Hamiltonian for the interaction between the ion and the applied laser field  is given by \index{electric dipole interaction}
\begin{eqnarray}
H^{(i)}
&= - {\bf E} \cdot {\bf D} = E \hat{\bf x} \cos (\omega_L t-kz )\cdot {\bf D} \nonumber
\\ &= - \frac{E}{2} \{ e^{i(\omega_L t - kz)} + e^{-i(\omega_L t -kz)}\}\hat{\bf x} \cdot {\bf D}
\end{eqnarray}
which describes the laser propagating along the $z$ direction polarised along $x$ and ${\bf D}$ is the atomic dipole operator.

\noindent
Now an operator $O$ can be written
\beq
O= \sum_{i,j}O_{ij} \vert i \rangle \langle j \vert
\label{eq:secondquantise}
\eeq
where $O_{ij}=\langle i \vert O \vert j\rangle$.

\noindent
So the operator $\hat{\bf x}\cdot {\bf D}$ can be written
\beq
\hat{\bf x}\cdot {\bf D} = \sum_{i,j} \langle i \vert \hat{\bf x}\cdot {\bf D}\vert j \rangle \vert i\rangle \langle j \vert = \langle e \vert \hat{\bf x}\cdot {\bf D}\vert g \rangle
\{
\vert e \rangle \langle g \vert + \vert g \rangle \langle e \vert
\}
\eeq

\noindent
This gives
\beq
H^{(i)} = \frac{\hbar \Omega}{2} (\sigma_+ + \sigma_-)
\{
 e^{i(\omega_L t - kz)} + e^{-i(\omega_L t -kz)}
\}
\eeq
\noindent
where we have introduced the definitions  $\sigma_+ = \vert e \rangle \langle g \vert$ and $\sigma_- = \vert g \rangle \langle e \vert$ and we define the Rabi frequency \index{Rabi frequency} $\Omega$
\beq
\Omega=-\frac{E}{\hbar}\langle e \vert\hat{\bf x}\cdot {\bf D} \vert g \rangle
\eeq

\noindent
Now $z=z_0 (a + a^\dagger)$ ($z_0=\sqrt{\hbar / 2m\nu}$ is the root mean square extension of the ground state wavefunction), and we define the \index{Lamb-Dicke parameter} Lamb-Dicke parameter $\eta=k z_0 $ giving
\beq
H^{(i)} = \frac{\hbar \Omega}{2} (\sigma_+ + \sigma_-)
\{
e^{i\eta (a+a^\dagger)} e^{-i\omega_L t} + e^{-i\eta (a+a^\dagger)} e^{i\omega_L t}
\}
\eeq
We will return later to an interpretation of the important Lamb-Dicke parameter (LDP). For now we consider the physical interpretation of this Hamiltonian. The Laser field can have two notable effects: it can (i) affect the motion of the active electron in the ion ($\sigma_+, \sigma_- $)  and it can (ii) affect the motion of the ion in the trap ($a,a^\dagger$). The first of these actions is fairly obvious since the laser radiation can be in resonance with the atomic transition. The second action is less obvious since the motional frequency is very low so that direct resonant coupling of this motion to the laser field is out of the question. Critically, it transpires that (i) and (ii) go hand in hand and the key to understanding the interaction is to see how this works.

This is best done by transforming to the `interaction picture'\index{interaction picture}. This is a useful step when a Hamiltonian can be separated into time independent parts and a part which contains all the explicit time dependence.
Looking back at equation \ref{fullham} we note that  $H_A$ and $H_m$ are themselves time independent and so lead only to the usual exponential time dependence associated with their respective eigenfunctions. On the other hand $H^{(i)}$ is itself a time dependent operator. The benefit of transforming to the `interaction picture' is that it allows us to home in on the effects caused by the time dependent interaction without the clutter of the normal time dependence associated with $H_A$ and $H_m$.

In general if we can write a Hamiltonian as $H=H_0+V(t)$ then $H$ obeys the Schr\"odinger equation
\beq
H\Psi=i\hbar \frac{\partial \Psi}{\partial t}
\eeq

Now if we define a new wavefunction $\Psi^\prime= e^{iH_0t/\hbar} \Psi$ then we also have $\Psi= e^{-iH_0t/\hbar} \Psi^\prime$. It can be shown that the full Hamiltonian in the interaction picture, denoted $\bar H$, is given by
\beq
\bar H= e^{iH_0t/\hbar} V e^{-iH_0t/\hbar}
\eeq
where this Hamiltonian obeys the Schr\" odinger equation
\beq
i \hbar\frac{\partial \Psi^\prime}{\partial t}
= \bar H \Psi^\prime
\eeq

Substituting $H_0=H_A+H_m$ into this equation leads (after some algebra involving Taylor expansions of exponentiated operators) and after making the rotating wave approximation (neglecting terms that evolve with frequency $\omega_L +\omega_a$) to an expression for the full Hamiltonian for our system in the interaction picture \index{Rabi frequency}. To simplify the notation we define $\tilde a=ae^{-i\nu t}$,  $\tilde a^\dagger =a^\dagger e^{i\nu t}$, and $\Delta=\omega_L-\omega_a$ giving

\beq
\bar H =\frac{\hbar\Omega}{2}
\{
e^{i\eta (\tilde a+\tilde a^\dagger)}\sigma_+ e^{-i\Delta t}
\} + H. c.
\label{eq:hamiltonian_int_final}
\eeq

Below we will see how  the experimentally adjustable detuning $\Delta$ plays a decisive role in determining the response of the trapped ion's motion and internal state to the applied radiation field. In what follows it is shown that choosing the right detuning picks out specific changes to the motional state to go hand in hand with the excitation or de-excitation of the internal state.

\subsection{Lamb-Dicke Regime}
\index{Lamb-Dicke regime}
\index{motional sidebands} The motion of the ion in the trap means that, in its rest frame, it sees the monochromatic laser as being comprised of a range of frequencies. The fact that the motion is simple harmonic means that only certain discrete frequencies are present in the spectrum seen by the ion in its rest frame. For large amplitude oscillations the spectrum consists of an unshifted (carrier) frequency plus an array of positive (blue) and negative (red) sidebands, separated by the motional frequency $\nu$ (the amplitude of the motion acts as a `modulation index' in FM theory). For small oscillations the spectrum becomes simpler with just the carrier and a single pair of sidebands (one red and one blue) having an appreciable amplitude. We will put these notions on a firm mathematical footing later. For now we note that we are free to choose the laser frequency we apply. If we choose to apply the laser with $\omega_L=\omega_a$ then the ion in its rest frame will interact with the carrier. On the other hand if we apply radiation that is detuned by the motional frequency $\omega_L=\omega_a\pm \omega$ then the ion, in its rest frame, will interact with one of the sidebands.

Returning to the lab frame we should ask how it is that a {\em detuned} photon can be absorbed by the ion, changing its internal state by more (or less) than the photon energy, and yet conserving energy. The answer is that the ion's motion can be de-excited (or excited) in the process restoring the energy balance. On the other hand, if we tune the laser to resonance then there is no change to the ion's motional energy as a result of the photon absorption.

We will restrict ourselves to the situation where the motion of the ion is small enough that a single pair of sidebands is present. This is known as the Lamb-Dicke regime. For typical miniature RF traps, Doppler cooling will put an ion into the Lamb-Dicke regime.
In the Lamb-Dicke regime the laser can cause a limited range of interactions: (i) it couples $\vert g,n\rangle$ to $\vert e, n\rangle$ by choosing it to be resonant or (ii) it couples $\vert g,n \rangle$ to $\vert e, n+ 1\rangle$ by detuning it to the blue sideband or (iii) it couples  $\vert g,n \rangle$ to $\vert e, n - 1\rangle$ by detuning it to the red sideband. We are therefore choosing $\Delta=\omega_L-\omega_a= \delta \pm\nu$ where $\delta$ allows for the un-ideal situation where the laser isn't exactly tuned to a sideband. We will show below how, mathematically, the choice of the detuning picks out the relevant interaction.

\index{Lamb-Dicke regime}
\index{motional sidebands} How small should the oscillations be in order for the spectrum to simplify to a carrier plus two sidebands? Classically the amplitude of the oscillation needs to be much smaller than $\lambda/2\pi$. The QM analogue of the amplitude of the oscillation is given by the spread of the wavefunction $\hat z_n$,  which is given by $(\langle z^2 \rangle)^{1/2}$ where
\beq
\langle z^2 \rangle=(2\langle n \rangle+1)\hbar/2m\nu
\eeq
 So that

\beq
\eta\sqrt{2\langle n \rangle+1}= \hat z_n\frac{2\pi}{\lambda}
\eeq

The classical Lamb-Dicke criterion says that the r.h.s. must be much smaller than 1.
 Starting with the Hamiltonian \ref{eq:hamiltonian_int_final}, below we will expand the operator with $a$ and $a^\dagger$ in the exponent.  Such an expansion to lowest order in $\eta$ adequately describes the ion's dynamics only if
\beq
\eta \sqrt{2\langle n \rangle +1} \ll 1
\eeq
which defines the QM Lamb-Dicke regime.

At this stage it is useful to consider a Taylor expansion of the operator $e^{i\eta(\tilde a + \tilde a^\dagger)}$. In the Lamb-Dicke regime we can limit the Taylor expansion to first order
\beq
e^{i\eta(\tilde a + \tilde a^\dagger)}= 1+i\eta(ae^{-i\nu t}+ a^\dagger e^{i\nu t})
\label{eq:first_order}
\eeq
The full Hamiltonian of equation~\ref{eq:hamiltonian_int_final} is completely general, but in the Lamb-Dicke regime, and given a particular laser tuning, it is well approximated by a much simpler Hamiltonian. In the same way as the rotating wave approximation picks out certain terms on account of their relatively slow time dependence (`stationary terms') the combination of $\Delta$ and $\nu$ in the above equation picks out particular terms when equation \ref{eq:first_order} is inserted in the full Hamiltonian \ref{eq:hamiltonian_int_final}.

For $\Delta=0$  (the carrier) and using equation~\ref{eq:first_order} we find
\beq
\bar H=\frac{\hbar \Omega}{2}(\sigma_+ + \sigma_-)
\eeq

For $\Delta=\nu$ (blue sideband) we have
\beq
\bar H=\frac{\hbar \Omega}{2}\eta(a^\dagger\sigma_+  +  a\sigma_-)
\label{eq:blue_sb}
\eeq

For $\Delta=-\nu$ (red sideband) we have
\beq
\bar H=\frac{\hbar \Omega}{2}\eta(a\sigma_+  +  a^\dagger\sigma_-)
\label{eq:red_sb}
\eeq

\index{Jaynes-Cummings Hamiltonian}
This takes exactly the same form as the Jaynes-Cummings Hamiltonian of quantum optics with the exception that here $a$ and $a^\dagger$ destroy and create quanta of motion in the harmonic well rather than photons of the EM field (the quanta of the motional state of an ion are often called phonons following the nomenclature used in condensed matter physics). We can read it as meaning that a phonon is destroyed when the ion is promoted from the $\vert g \rangle$ to $\vert e \rangle$ and a phonon is created when the ion makes a transition from $\vert e \rangle$ to $\vert g \rangle$. For this reason the Hamiltonian for the blue sideband is often referred to as the an anti-Jaynes Cummings Hamiltonian which is a new possibility opened up in  this context.

\subsection{Coupling Strength}

\index{Rabi frequency}
The strength of the coupling between an atom and the radiation field is given by the Rabi frequency.
Building upon the ideas presented in the last section is is straightforward to calculate the coupling strength for interactions when the laser is detuned to the sidebands. This is taken into account through a modified Rabi frequency $\Omega_{m+n,n}$. In the Lamb-Dicke regime only three interactions need to be considered, one for the carrier

\beq
\Omega_{n,n} = \Omega (1-\eta^2 n)
\eeq

and one for each of the two sidebands

\beq
\Omega_{n+1,n} = \eta\sqrt{n+1}\ \Omega\ \ \
{\rm blue\ sideband}
\eeq

\beq
\Omega_{n-1,n} = \eta\sqrt{n}\ \Omega\ \ \
{\rm red\ sideband}
\eeq

\index{motional sidebands} For small $\eta$, these relations demonstrate the expected result that interaction on the sidebands is weaker than that on the carrier. In classical FM theory the red and blue sidebands have equal weight however note here there is a purely quantum mechanical effect. For a given $n$ the coupling on the red sideband is a little weaker than on the blue sideband.  This imbalance gets greater as the system pushes deeper into the quantum regime where $n$ is small.  When $n=0$ the coupling on the red sideband is zero. This is exactly what one would expect -- with the laser tuned to the position of the red sideband and the ion already in $\vert n=0\rangle$ there is no motional state below $\vert n=0\rangle$ to which the excitation can take place. By tuning the laser to the red sideband the effect on the motional state is therefore {\em conditional} upon the initial motional state of the ion. If it is in an excited motional state then an interaction can occur whereas if it is already in the motional ground state nothing happens. This conditionality was exploited by Cirac and Zoller \cite{ciraczoller} in their original proposal for a two-qubit quantum gate using trapped ions (see figure ~\ref{Sideband_Cool}).

It is worth noting that while small $\eta$ is desirable, since it simplifies the interactions, if $\eta$ is {\em too}
 small then the interactions on the sidebands will be very weak. For applications such as normal sideband cooling described in the next section it is therefore necessary to use visible radiation since microwave transitions lead to very small values of $\eta$ and so do not allow for simultaneous changes to the motional state of an ion that go hand in hand with changes to the internal state. Another way of viewing this is in terms of the momentum $\hbar k$ of the photons involved. For small $k$, as found in the microwave region of the EM spectrum, the photons simply do not carry enough momentum to give the ion the required momentum kick in order to change its motional state.
 
\subsection{Sideband Cooling}
\label{sec:SBCoolingLaser}
\begin{figure}
\centerline{\psfig{file=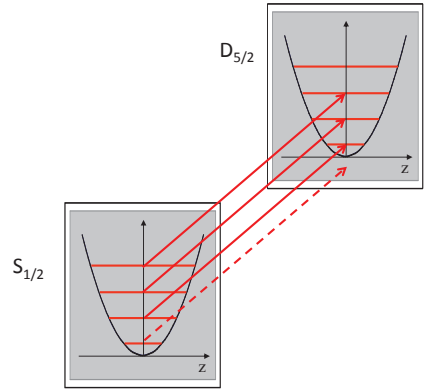,width=9cm}}
\caption{The state of the ion is determined by its internal state (S$_{1/2}$ or D$_{5/2}$ for Ca$^+$) and its quantum state of motion in the trap potential. By tuning a laser to the red sideband transitions that result in the loss of one quantum of motional energy are excited. When the ion returns to the lower internal state it does so predominately conserving the motional quantum number. The ion thus steps down the ladder of motional states until it is in the ground state. Once the ground state is reached the interaction with the laser automatically switches off since there is no state below the ground state to which excitation can occur.}
\label{Sideband_Cool}
\end{figure}

\index{sideband cooling}
Imagine a single ion that has already been cooled on a strongly allowed transition to the Doppler limit in a trap which is sufficiently stiff that the ion is in the Lamb-Dicke regime with small $\eta\lessapprox 1$. To remove the last few quanta of motional energy we switch to `sideband cooling' on a narrow transition. By tuning the laser to the red sideband a photon will be absorbed which excites the ion internally but removes a quantum of motional energy (see figure\ref{Sideband_Cool}). The fact that the ion is in the Lamb-Dicke regime means that in emission the carrier transition is favoured so that when the ion returns to the internal ground state the motional state is on average unaltered. Just a few cycles of this process should leave the ion in the motional ground state. Once the ion is in the motional ground state the system uncouples from the radiation since there is no motional state below the ground state to which a transition might occur (as described above).

As described so far, this scheme has a major drawback. The cycle time is set by the natural lifetime of the upper state which is necessarily very long (since it is associated with a narrow transition). For example with  Ca$^+$, it might take a second or longer for the ion to complete a cycle, if the upper state is metastable. Heating rates in small traps are usually not insignificant at this level so that some method for achieving a modest speed-up of the process is needed. This can be achieved by using another laser to couple the upper state of the narrow transition via an allowed transition to a higher lying state which {\em does} have a dipole-allowed transition to the internal ground state of the ion (for Ca$^+$ this is the $D_{5/2} \rightarrow P_{3/2}$ transition at 854nm, see figure~\ref{Ca+LevelScheme}). The effect is to `lend some width' to the narrow transition so that the ion recycles to the ground state more rapidly \cite{Hendricks2008}.

An intercombination resonance in In$^+$ is sufficiently narrow (natural linewidth $\Gamma = 2\pi\times 360$ kHz) to be able to resolve motional sideband in typical ion traps, but still provides fast spontaneous decay to close the sideband cooling cycle. Sideband cooling of In$^+$ is reported in \cite{Peik1999}.

There are some practical limits to the efficacy of sideband cooling. One of these is associated with an over-simplification we have made in the discussion of the process. If a laser is tuned to the red sideband it is still capable of resulting in `off resonant' excitation on the carrier or, worse, on the blue sideband. Though these latter excitations will be rare, they actually put energy {\em into} the ion motion. Schemes for overcoming this drawback have been developed (see EIT cooling below).

We have assumed that one privileged direction in the ion trap has a lower oscillation frequency than the others and have discussed sideband cooling in the context of the motion in this one direction. If cooling to the motional ground state in 3D is required, then the process needs to be stepped through the different frequencies of motion, adding a complication. If coupling between the motion in different directions is weak (as is usually the case) it may suffice to perform ground state cooling in only one dimension leaving the ions relatively hot in the other modes (directions).

Sideband cooling was first performed for a single Hg$^+$ ion \cite{Diedrich1989}. It has since been performed for a number of other species of ion  (e.g.  \cite{Peik1999, Letchumanan2007, Schmidt-Kaler2000, Schwedes2004}).  It has also been achieved for short strings of ions in radiofrequency traps \cite{King1998, Monz2011}. An added complication for strings of ions is the additional motional modes of oscillation. For a single ion there is only one mode of oscillation in 1D -- the centre of mass (COM) mode. For two ions, there are two modes -- the COM mode in which the two ions move in phase with each other and the `breathing mode' in which two ions move in anti-phase. Each extra ion added to a string of ions in the trap brings an extra mode in 1D. In reality the spectrum of even a short string of ions in a trap is significantly more complicated than that alluded to in the discussion above of a carrier and a single pair of sidebands.

\subsubsection{Raman sideband excitation}
\index{Raman sideband cooling}
In the sections above it became clear why, for sideband cooling, laser light was used. The Lamb-Dicke parameter $\eta$, that is a measure for how well the internal dynamics of trapped ions can be coupled to their vibrational dynamics, should not be too small in order to be able to efficiently excite a motional sideband for cooling. Thus, typical Zeeman and hyperfine resonances in the RF regime (of order 1 MHz - 10 GHz) would not be suitable for sideband cooling. However, if a two-photon Raman  transition between two such states is driven, then cooling becomes possible. The physical mechanism behind Raman sideband cooling will be outlined below.

If the two light beams driving a stimulated Raman transition between hyperfine states were parallel, that is,  $\hat{k_1} = \hat{k_2}$, where $\hat{k_i} \equiv \vec{k_i}/|\vec{k_i}|, i=1,2$ is a unit vector pointing along the propagation direction of the respective light beam, then the absorption of a photon from one beam and subsequent stimulated emission into the second Raman beam would lead to a negligible net transfer of momentum to the atom. The momentum transferred to the atom, potentially useful for cooling, would amount to $\hbar\vec{k_1}-\hbar\vec{k_2}$, where the photons have a frequency difference corresponding to a typical  Zeeman or hyperfine splitting. This is, in typical traps, not sufficient to efficiently (de-)excite vibrational motion. However, if the two Raman beams are antiparallel (i.e., $\hat{k_1} = -\hat{k_2}$), then the net momentum transfer is $\hbar\vec{k_1}+\hbar\vec{k_2}$ making (de-)excitation of vibrational motion efficient \cite{Monroe1995,Morigi2001}.  The difference vector $\hat{k_1}  -\hat{k_2}$ points along the direction of the vibrational motion to be cooled, for example, an axial or radial vibrational mode in a linear ion trap.

\subsection{Sideband Cooling using RF radiation}
\label{sec:mw_cooling}
\index{sideband cooling using RF radiation}
Above, we saw that Zeeman or hyperfine resonances of the electronic ground state can be used for Raman sideband cooling. Driving such a resonance in 1-photon absorption and emission using RF radiation\footnote{Here we adhere to the definition of RF according to the Oxford English Dictionary: frequencies between about 300 GHz and 3 kHz } results in negligible momentum transfer to the ion and thus does not allow for efficient cooling. For example, the Lamb-Dicke parameter $\eta \approx 10^{-7}$ for the ground state hyperfine resonance in \yb.  The following paragraphs are devoted to outlining a scheme that allows for controlling the   motion of trapped ions using RF radiation, despite $\eta$ having a negligible magnitude.

\begin{figure}
\centerline{\psfig{file=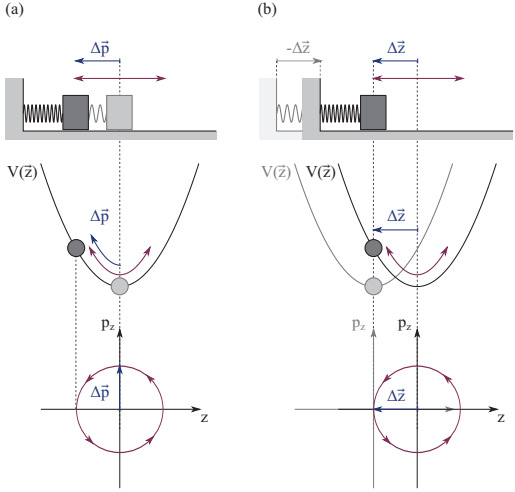,width=7cm}}
\caption{Two ways of exciting  a classical harmonic oscillator. a) A momentum kick $\Delta p$ sets the mass $m$ in motion. The trajectory in phase space is a circle. b) The oscillator's equilibrium position is nearly instantaneously shifted to the right by an amount $\Delta z$ resulting again in harmonic motion of $m$ around the new (shifted by $\Delta z$) equilibrium position.}
\label{fig:LDP_Classic}
\end{figure}

\index{sideband cooling using RF radiation, classical analogy} An illustration from classical physics may be useful in order to better understand how the ionic oscillator may be excited even though essentially no linear momentum is transferred from the absorbed photon. As a prototype of an harmonic oscillator characterized by angular frequency $\nu$ we consider a mass $m$ attached to a massless spring exerting a force $\vec{F_S}=-m\nu^2 \vec{z}$ on $m$ as sketched in Fig. \ref{fig:LDP_Classic}a. We start with a motionless oscillator. In order to set the mass in motion, we can apply a force $F$ for a time $\Delta t$ resulting in a momentum kick $ \Delta p=F/\Delta t$ and, as a consequence, the mass oscillates around its equilibrium position.  In phase space, the impulsive excitation corresponds to a shift along the momentum axis  followed by motion on a circle in phase space (Fig. \ref{fig:LDP_Classic}a). This motion is associated with cyclic transformation of the initial kinetic energy $ (\Delta p)^2/2m$ into potential energy and vice versa. This way of exciting an oscillator would correspond to the motional excitation of a trapped ion by absorption of photon associated with a momentum kick. In the case of a quantum mechanical oscillator, energy conservation is satisfied by appropriately tuning the light field to a sideband (see equations \ref{eq:blue_sb} and \ref{eq:red_sb}).

Alternatively, the suspension point of the spring may be (nearly) instantaneously moved to a new equilibrium position (i.e., on a timescale $\Delta t$ fast compared to the oscillation frequency of the mass), $\Delta z$ away from the initial position. Due to its inertia, the mass $m$ does not initially move. However, now it is no longer located at the equilibrium position of the harmonic potential (corresponding to the spring in its relaxed state). Instead, it is displaced from the spring's new equilibrium position, has acquired the  potential energy $m\nu^2(\Delta z)^2$, and thus will start to oscillate around this new equilibrium position. The corresponding trajectory in phase space is again a circle signifying harmonic motion. A quantum mechanical analogue to this type of excitation is discussed in the next paragraphs.

\begin{figure}
\centerline{\psfig{file=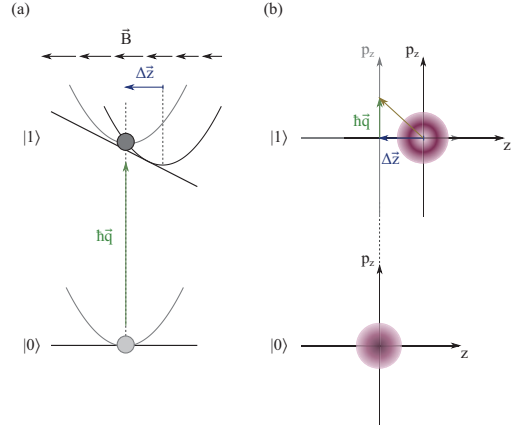,width=7cm}}
\caption{Excitation of a quantum mechanical  harmonic oscillator by a state-dependently shifting its equilibrium position. a) Upon excitation to state $\ket{1}$ the equilibrium position of the harmonic oscillator is shifted by an amount $\Delta z$. This shift is caused here by an additional linear potential superimposed with the harmonic oscillator potential. This energy shift may be due to a spatially varying magnetic field giving rise to a position dependent Zeeman shift of state $\ket{1}$. b) Illustration in phase space. Here, the oscillator is excited from its ground state to its first excited state (a Fock state characterized by vibrational quantum number $n=1$).  In addition, if the energy splitting between the atomic states, $\hbar \omega_0$is sufficiently large, the atom may experience an appreciable momentum kick $\hbar \omega_0 /c$ associated with the absorption of a photon. This is not the case if the resonance $\omega_0$ lies in the RF regime. }
\label{fig:LDP_QM}
\end{figure}

Adding a state dependent force \index{state dependent force} to the effective harmonic force confining a trapped ion has a similar effect. This is illustrated in Fig. \ref{fig:LDP_QM}. Upon excitation from state $\ket{0}$ into state $\ket{1}$ the oscillator's equilibrium position is shifted by
\be
\Delta z=F/(m\nu^2)
\label{eq:Deltaz}
\ee
where $F=|\partial_z E_1(z)|$ or, using $E_1=\hbar\omega$,
\be
F=(\hbar/2)|\partial_z \omega(z)|,
\ee
that is,  \index{magnetic gradient induced coupling} the force $F$ is proportional to the magnitude of the gradient $|\partial_z \omega(z)|$ of the ionic angular resonance frequency $\omega(z)$. Such a gradient can be achieved by applying a spatially varying magnetic field $B(z)$ to the trapped ion \cite{Mintert2001}. \index{Lamb-Dicke parameter, effective} The parameter that replaces the usual LDP and serves as a measure for how likely the motional excitation of an ion is when exciting its internal resonance is
\be
\kappa \equiv \Delta z/z_0
\label{eq:kappa}
\ee
with the root mean square extension of the ground state wavefunction, $z_0$  of a trapped ion. Using eqs. \ref{eq:Deltaz} and \ref{eq:kappa}, the coupling strength $\kappa$ can also be expressed as
\be
\kappa =\frac{ \ z_0 |\partial_z \omega(z)| }{  \nu}
\ee
which indicates that $\kappa$ measures the change in the ion's resonance frequency when the ion is moved over a distance $ z_0$ in units of the harmonic oscillator frequency.  Thus, coupling between internal and motional dynamics becomes efficient, and cooling by scattering of photons is possible,  even when RF radiation is used \cite{Mintert2001,Wunderlich2005,Johanning2009}.

A proper mathematical description of the interaction between a trapped atom and electromagnetic radiation in the RF regime shows that the the usual LDP $\eta$ is replaced by a new effective LDP \cite{Mintert2001},
\be
\eta_{\text{eff}} = \eta + i\kappa \equiv \eta' e^{i\theta}\  .
\label{eq:eta_eff}
\ee
Even if the motional excitation of the ion relies on the state dependent oscillator potential as described above, there is linear momentum associated with the absorption of a photon (i.e., LDP $\eta \neq 0$). However, this momentum is appreciable only if radiation of high enough frequency is employed to drive the ionic resonance. In eq. \ref{eq:eta_eff}, $\eta$ accounts for absorption of linear momentum while $\kappa$ accounts for the effect due to the gradient in the resonance frequency. This expression reduces to $\eta_{\text{eff}} = \kappa $ (neglecting a global phase factor) when $\eta \approx 0$, which is the case for RF radiation. Then, the Hamiltonian reads \cite{Mintert2001} (in an interaction picture under the RWA, expanded up to first order in $\eta_{\text{eff}}$
[compare eqs. \ref{eq:hamiltonian_int_final} and \ref{eq:first_order}])
\be
\tilde{H}_l=\frac{1}{2}\hbar\Omega_R\left[e^{-i\left[\Delta t+\phi\right]} \sigma_+  \left[1+i\eta_{\text{eff}}\left(a^{\dagger}e^{i\nu_z t}+ae^{-i\nu_z t}\right)
\right]+H.c.\right].
\ee
If we set the detuning $\Delta=-\nu_z$, then we obtain the Hamiltonian eq. \ref{eq:red_sb}
with $\eta$ replaced by $\eta_{\text{eff}}$, which is the Hamiltonian that was shown to make sideband cooling possible (the phase $\phi$ in the above Hamiltonian is not relevant here). Thus, in order to achieve cooling, the red motional sideband can be driven by long-wavelength RF radiation \cite{Khromova2012}. Such resonances in the electronic ground state are characterized by the absence of spontaneous emission (for all practical purposes).  Therefore, after absorption of a photon on the red sideband, and the ensuing loss of one motional quantum, the ion has to be returned to its initial state by a suitable optical pumping process as was mentioned in section~\ref{sec:SBCoolingLaser}.

The field that leads to \index{magnetic gradient induced coupling} magnetic gradient induced coupling (MAGIC) between internal and motional states of ions, at the same time, makes ions in an ion Coulomb crystal individually distinguishable by their resonance frequency \cite{Mintert2001,Johanning2009,Wang2009}. This is the case, if the ionic resonance exhibits a Zeeman shift. For this purpose, instead of applying a magnetic gradient one could also use a spatially varying ac Stark shift induced by an additional light beam \cite{Staanum2002} to individually address ions by selecting their respective resonance frequency. In addition, this Stark shift gradient  gives rise to an effective LDP
that could potentially be used to drive motional sidebands in the long-wavelength range. The magnitude of this effective LDP induced by an ac Stark shift gradient will depend on atomic properties that, for the typically used ion species,  are known or could be measured, and on the intensity profile of the light beam, and its detuning. Another possibility for sideband cooling is to take advantage of the gradient of a RF field driving a hyperfine transition  to induce coupling between internal and motional states and thus sideband cool trapped ions \cite{Ospelkaus2011}.

\subsubsection{Simultaneous cooling of many vibrational modes}
\index{simultaneous cooling of vibrational modes}
Usually, if one wants to sideband-cool $N$ modes  of an $N-$ion crystal (considering one spatial direction only), the radiation driving an internal atomic resonance is red-detuned by the frequency of the motional mode to be cooled. Thus, only this one mode looses energy during repeated cooling cycles and is cooled efficiently. If cooling of more than one mode is desired, then the light field has to be tuned sequentially to the resonances corresponding to the respective modes (or, alternatively,  light fields at multiple sideband resonances are applied simultaneously).
In a Coulomb crystal consisting of $N$ trapped ions, the magnetic gradient that induces $\eta_{\text{eff}} $, at the same time, shifts the resonance frequency of each ion depending on its position in the crystal. This effect can be put to use for  simultaneously cooling a number of vibrational modes \cite{Wunderlich2005} as is outlined below.

The magnetic gradient could be designed so that the red sideband resonance, corresponding to the COM mode at frequency $\nu_1$, of  ion 1 coincides with the red sideband, corresponding to the mode with the next higher frequency $\nu_2$, of ion 2, and so on for the other ions . In this way, a red sideband resonance of each one of $N$ ions is used for cooling a particular vibrational mode. Since all these resonances coincide, it is sufficient to irradiate all ions with just a single frequency field in order to achieve simultaneous cooling of all $N$ vibrational modes.

\section{Laser cooling using electromagnetically induced transparency}
\index{electromagnetically induced transparency}
We have seen how simultaneous cooling of several vibrational modes could be achieved by suitably shaping the atomic absorption spectrum such that
several motional red sideband are simultaneously excited with a fixed-frequency field.  A suitably shaped atomic spectrum that allows for simultaneous cooling of more than one mode may also be achieved by employing electromagnetically induced transparency.

As mentioned in section\ref{sec:SBCoolingLaser}, when applying usual \index{sideband cooling} sideband cooling, a limit on the achievable cooling rate is imposed by power broadening of the red sideband resonance which leads to non-resonant excitation of the carrier resonance and even of the blue sideband, thus giving rise to transitions that contribute to heating of the ions. If these spurious transitions could be suppressed, then the cooling limit and the cooling rate would benefit.  Therefore, a second effect of a deliberately and suitably modified atomic absorption spectrum (in addition to the cooling of multiple modes) useful for cooling can be that it effectively suppresses unwanted resonances contributing to heating of the ions, for example, the carrier and blue sideband resonance.

\begin{figure}
\centerline{\psfig{file=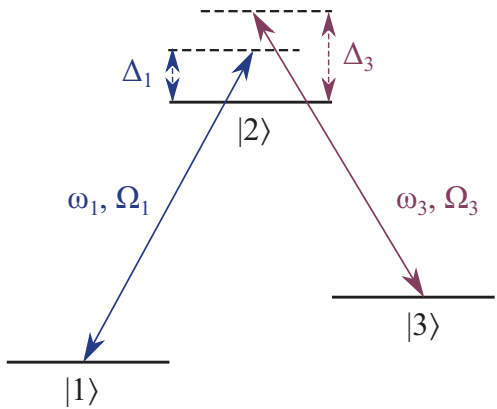,width=7cm}}
\caption{Generic 3-level system that exhibits electromagnetically induced transparency useful for cooling of trapped ions. }
\label{fig:ThreeLevel}
\end{figure}

We will now outline how the atomic spectrum can be modified in a desired way using coherent driving fields. For our purposes, the latter typically are in the RF range, or between the infra-red and ultra-violet range of the electromagnetic spectrum. For brevity, we will refer to such an electromagnetic field  as the ``light field'' or simply ``field''.
Different theoretical treatments of the light-atom interaction may be used here. A perturbative treatment is valid only under particular conditions imposed on the relative strength of the parameters characterizing the atom interacting with the field(s). The optical \index{optical Bloch equations} Bloch equations are exact but do not always provide clear insight into the physical processes that take place in the atom-field interaction. \index{dressed states} The dressed state picture can be useful to identify states and scattering amplitudes that play a role for a given set of parameters. Here, we view the atom-field interaction from each one of these three different angles depending on the context and on the insight we wish to obtain.

For the moment we are interested in coherent dynamics only, and do not take into account spontaneous emission.
If we deal with hyperfine and Zeeman states connected via magnetic dipole resonances in the RF regime, then this is a valid approach. However, for laser cooling an additional dissipative process is required. If a three-level system as shown in Fig. \ref{fig:ThreeLevel} is implemented with optical resonances, this requirement is realized by dipole allowed spontaneous emission from state $\ket{2}$ into states $\ket{1}$ and $\ket{3}$ with rates $\beta\times \Gamma$ and $(1-\beta)\times \Gamma$, respectively ($\beta \in [0,1]$). Incidentally, this implies that there is no dipole allowed transition between states $\ket{1}$ and $\ket{3}$. Long-lived hyperfine ground states may be coupled to an extra level by laser light to induce an effective spontaneous decay of one of these states.

We consider a three-level atom and two light fields as depicted in Fig. \ref{fig:ThreeLevel}. First, it will be shown how the atomic response to a probe field (here, at frequency $\omega_3$ ) can be modified when applying a second driving field (here, at frequency $\omega_1$ ). For explaining how the spectral response of an atom can be modified using coherent light, the dressed atom picture is useful \cite{CCT1966,CCT1977,CCT2004} where the electromagnetic field quantized.

In order to appreciate how the absorption spectrum is modified, let us first consider the atomic system reduced to two levels $\{\ket{1},\ket{2}\}$ driven by a  field with frequency $\omega_{1}$. The dipole interaction between the field $\vec{F_1}\cos(\omega_1 t)$ with (electric or magnetic) amplitude $\vec{F_1}$ field and the atomic (electric or magnetic) dipole $\vec{d_1}$ is
\be
V = -\vec{d_1}\cdot\vec{F_1}/\hbar \propto (b_1+ b^\dagger_1)
\ee
 ($b^\dagger_1$ and $b_1$ are the creation and annihilation operator, respectively for the field mode at frequency $\omega_1$). For a linearly polarized quantized field whose axis of polarization is adopted as the $x-$axis, this interaction reads (using $\sigma_x=\sigma_+ + \sigma_-$, not considering a time dependence)
\be
V   &=& \frac{\hbar\Omega_1}{2}  \left[(\sigma_+ b_1 + \sigma_- b^\dagger_1 ) + (\sigma_+ b^\dagger_1  + \sigma_- b_1)  \right]  \ .
\label{eq:V}
\ee
The \index{Rabi frequency} Rabi frequency $\Omega_1$ (here chosen to be a real number) characterizes the strength of the coupling. Below, we will retain only the first two terms on the r.h.s. of  equation  \ref{eq:V} which describe resonant coupling between atom and field (corresponding to the rotating wave approximation in a semi-classical description). These terms describe the excitation of the atom while simultaneously taking away one photon from the field, or vice versa.

\begin{figure}[h]
\begin{center}
  \parbox{2.1in}{\epsfig{figure=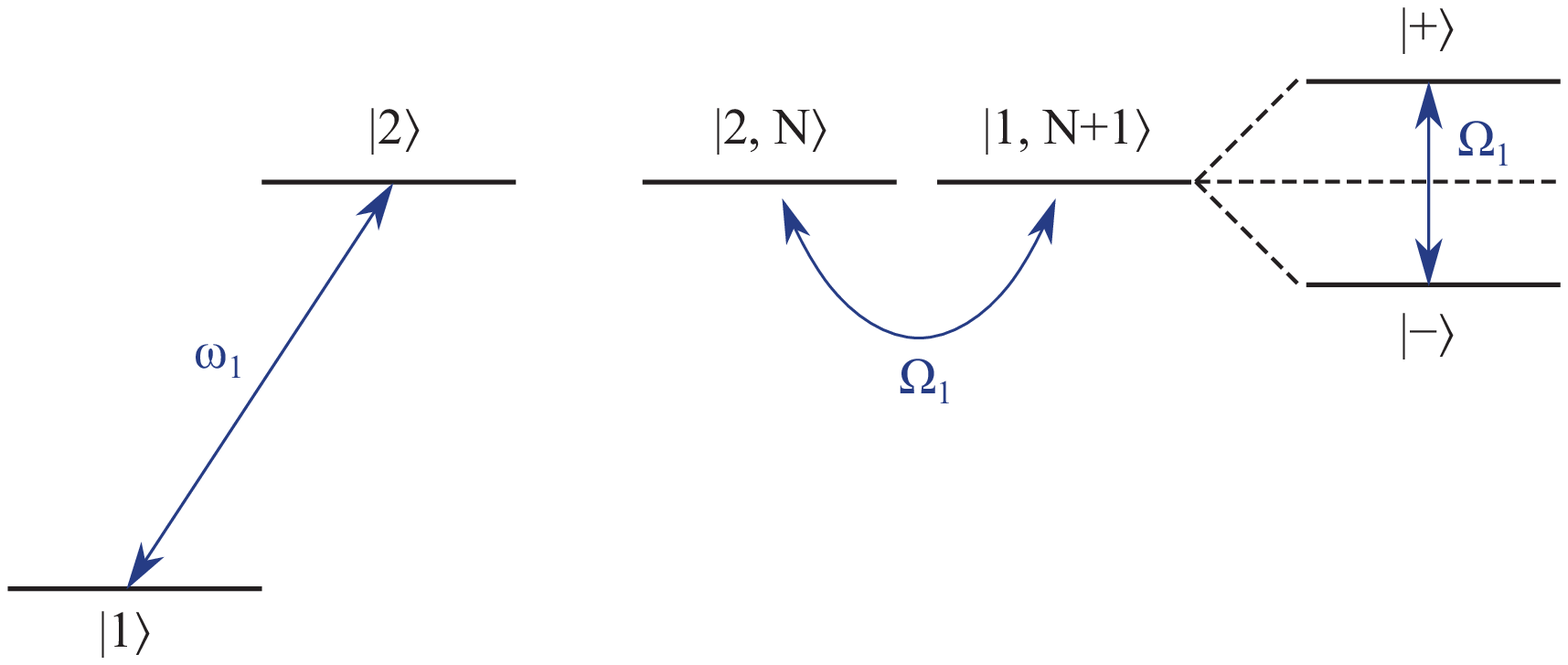,width=2in}}
  \hspace*{4pt}
  \parbox{2.1in}{\epsfig{figure=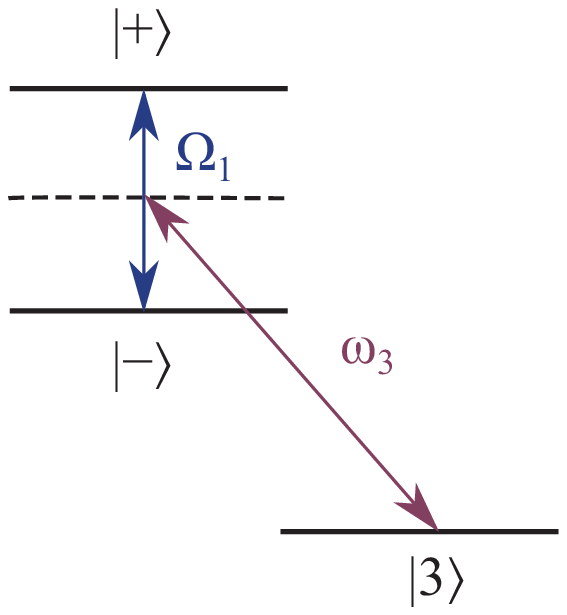,width=2in}}
  \caption{
  (a) Atomic states $\{\ket{1},\ket{2}\}$ dressed by the field with frequency $\omega_1$. When the field is included explicitly into the Hamiltonian, then state $\ket{1}$ with $N+1$ photons is degenerate with state $\ket{2}$ in the presence of $N$ photons. The dipole coupling between field and atom characterized by Rabi frequency $\Omega_1$ lifts this degeneracy and gives rise to the eigenstates (dressed states) $\ket{+}$ and $\ket{-}$.
  (b) The weak probe field at frequency $\omega_3$ is swept across the the level structure formed by the dressed states $\ket{+}$ and $\ket{-}$ (here, the probe field is tuned midway between the two expected resonances).  }
  \label{fig:DressedStates}
\end{center}
\end{figure}

Thus,  the Hamiltonian describing the atom-field system is
\begin{equation}
H_1 = \frac{\hbar\omega_a}{2}\sigma_z  + \hbar\omega_1 b^\dagger_1 b_1
        + \frac{\hbar\Omega_1}{2}(\sigma_+b_1 + \sigma_- b_1^\dagger) \  ,
\label{eq:H_1}
\end{equation}
where we have included the field itself, not only the dipole interaction with the atom, into the Hamiltonian.
The first two terms on the r.h.s. of eq. \ref{eq:H_1} describe the eigenenergies of the uncoupled atom-field system. In Fig. \ref{fig:DressedStates}a) these energy levels are sketched for the resonant case (i.e., $\Delta_1 =\omega_a-\omega_1=0$): The atomic level $\ket{1}$ in the presence of $N+1$ photons is degenerate with atomic level $\ket{2}$ in the presence of $N$ photons.

When the coupling term (last term in eq. \ref{eq:H_1}) is taken into account, then the field $\omega_1$ gives rise to the \index{dressed states} dressed states ($\Delta_1=0$, Fig. \ref{fig:DressedStates}a)
\be
\ket{+,N}&=& 1/\sqrt{2}(\ket{1,N+1}+\ket{2,N})   \\
\ket{-,N}&=& 1/\sqrt{2}(\ket{1,N+1}-\ket{2,N})
\label{equal}
\ee
where $N$ is the number of photons at frequency $\omega_1$ present in the field (Fig. \ref{fig:DressedStates}a).
These are stationary eigenstates of $H_1$ (eq. \ref{eq:H_1}).
Although we use the dressed state picture that arises when the field is quantized, here we don't take into account particular effects due to field quantization (the ``graininess'' of the field).  Instead, we consider strong coherent fields with the field amplitude proportional to $\sqrt{\bar{N}}$, and with a large mean photon number $\bar{N}$ (i.e.,  small relative fluctuations $\Delta N/\bar{N} \ll 1$,  but $\Delta N \gg 1$).
We are not interested in small changes to the field due to the presence of the atom, and, therefore, omit the label $N$ henceforth.

Obviously, atomic motion is not included in the Hamiltonian \ref{eq:H_1} (recall that $b^\dagger$ and $b$ refer to the electromagnetic field). At this point we are interested only in understanding how the atomic absorption profile can be modified.

\begin{figure}[h]
\begin{center}
  \parbox{2.1in}{\epsfig{figure=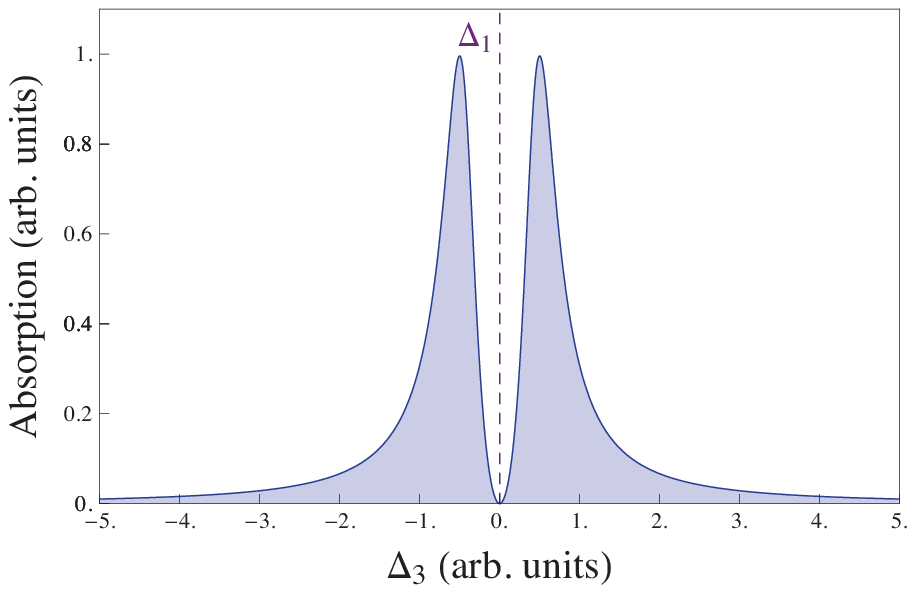,width=2in}}
  \hspace*{4pt}
  \parbox{2.1in}{\epsfig{figure=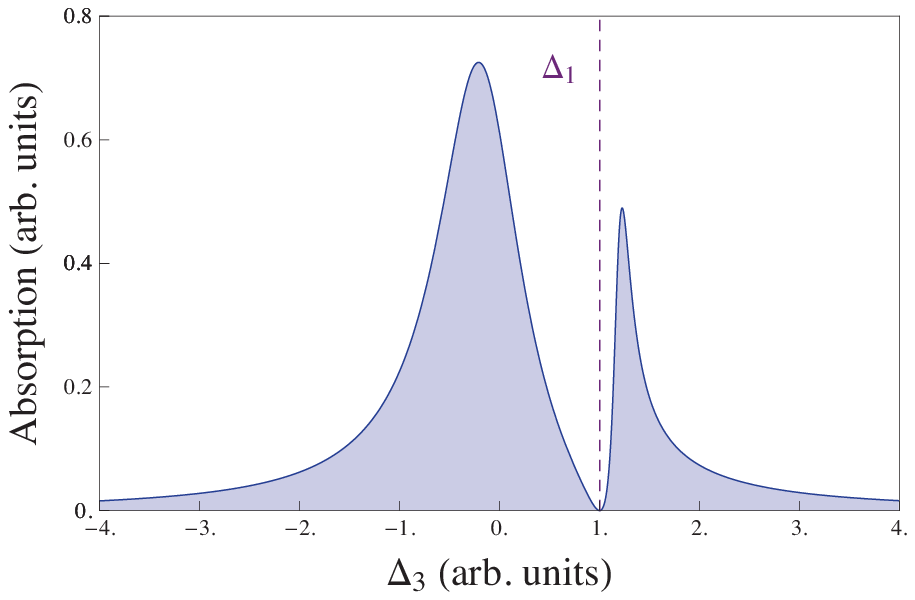,width=2in}}
  \caption{
Calculated absorption spectra of the 3-level system shown in Figs. \ref{fig:ThreeLevel} and \ref{fig:DressedStates} as a function of the detuning $\Delta_3$ of the probe field.
(a) The field dressing states $\ket{1}$ and $\ket{2}$ is tuned to exact resonance ($\Delta_1=0$). Two absorption maxima are observed  while zero absorption occurs when the probe field is tuned midway between the two dressed states $\ket{+}$ and $\ket{-}$ (compare Fig. \ref{fig:DressedStates}b), that is, $\Delta_3=0$.  Parameters (dimensionless, in units of the decay rate $\Gamma$ of state $\ket{2}$): $\beta=0.5$, $\Omega_1=1$, $\Omega_3=0.01$, $\Delta_1=0$.
(b) Detuning $\Delta_1 > 0$. The zero of the absorption probability occurs again at the two-photon-resonance condition $\Delta_3= \Delta_1$.  Parameters (dimensionless, in units of $\Gamma$):  $\beta=0.5$, $\Omega_1=1$, $\Omega_3=0.2$, $\Delta_1=1$.
   }
  \label{fig:EIT}
\end{center}
\end{figure}

The eigenstates of the atom-field system $\ket{+}$ and $\ket{-}$ are split by $\hbar\Omega_1$ (Fig. \ref{fig:DressedStates}). Now, we add the weak probe field with frequency  $\omega_{3}$ and Rabi frequency $\Omega_3 = -\vec{d_3}\cdot\vec{F_3}/\hbar$ (see Fig. \ref{fig:ThreeLevel}) scanning the level structure induced by the first field $\omega_1$. For now, we assume $\Omega_3 \ll \Omega_1$ so that the probe field does not appreciably alter the level structure \footnote{This assumption is introduced only for  ``pedagogical'' reasons. It is not necessary for explaining the absorption spectra and will be dropped later.}.

From the level structure displayed in Fig. \ref{fig:DressedStates} we expect to obtain two resonances, when monitoring the absorption of this probe field as a function of its detuning,
$\Delta_3$. This is indeed the case as is shown in Fig. \ref{fig:EIT}a. The width of the resonances seen in Fig. \ref{fig:EIT} is determined by the spontaneous decay of state $\ket{2}$ with rate $(1-\beta)\Gamma$ into state $\ket{1}$ and with rate $\beta\times\Gamma$ into state $\ket{3}$.
The fact that the absorption probability goes to zero for $\Delta_3=0=\Delta_1$ can be understood by considering the amplitudes for dipole transitions between state $\ket{3}$ and the states dressed by the $\omega_1$-field, $\ket{+}$ and $\ket{-}$, respectively. These transition amplitudes are (in the perturbative limit regarding the interaction between field 3 and the atom) proportional to
\be
\frac{\bra{+}\vec{d}\ket{3}}{\Omega_1/2 - \Delta_3 + i\Gamma/2 }
\ee
and
\be
 \frac{\bra{-}\vec{d}\ket{3}}{\Omega_1/2 + \Delta_3 + i\Gamma/2} \ ,
\ee
respectively. We note that
\be
\bra{+}\vec{d}\ket{3}&=&1/\sqrt{2}(\bra{1}+\bra{2})\vec{d}\ket{3}=+1/\sqrt{2}\bra{2}\vec{d}\ket{3}  \\
\bra{-}\vec{d}\ket{3}&=&-1/\sqrt{2}\bra{2}\vec{d}\ket{3}
\ee
(recall that there is no atomic dipole on the 1-3 resonance, $\bra{1}\vec{d}\ket{3}=0)$). Therefore, if the probe field is tuned midway between the two resonances ($\Delta_3=0$), both induced atomic dipole moments contribute equally in magnitude but with opposite phase and thus cancel exactly giving zero absorption probability. The atom does not scatter light at frequency $\omega_3$, in other words, it is transparent for this light, or viewed differently, the atomic sample remains dark.

Such a dark state was observed in sodium vapour \cite{Alzetta1976} and in numerous other experiments. A review of related theoretical and experimental work is found in references \cite{Arimondo1996,Fleischhauer2005}.  Dark resonances in ionic spectra are reported, for example,  using trapped Ba$^+$ ions \cite{Janik1985,Stalgies1996,Stalgies1998}, with Ca$^+$ \cite{Kurth1995,McDonnell2004,Lisowski2005,Albert2011}, with Sr$^+$ \cite{Barwood1998,Lindvall2012}, and Yb$^+$ \cite{Klein1990}, and have been used for cooling of Ba$^+$ \cite{Reiss2002} and Ca$^+$ \cite{Roos2000}.

A spectrum that displays a single resonance line for $\Omega_1 \rightarrow 0$ and two resonances for $\Omega_1 > 0$, as shown in Fig. \ref{fig:EIT}a), is also known as Autler-Townes doublet \cite{Autler1955}.

So far, in order to obtain some insight into the physical mechanism of EIT, we considered particular combinations of parameters of the light field and the atom ($\Delta_1=0, \Omega_3 \ll \Omega_1$). The description of the atomic response can, of course, be generalized by allowing for arbitrary detuning $\Delta_1$ and no longer requiring $\Omega_3 \ll \Omega_1$ when calculating the EIT absorption profile \cite{Harris1997,Fleischhauer2005}.

In the considerations above, the field at frequency $\omega_3$ was considered to be a weak field that probes the level structure induced by the field at frequency $\omega_1$. Now, when the coupling strengths  $\Omega_3$ and $ \Omega_1$ are comparable in size, the field at frequency $\omega_3$ itself alters the atomic response appreciably. A characteristic feature of the atomic spectrum remains: a dark resonance (no scattered light) arises when $\Delta_3=\Delta_1$
\cite{Harris1997}.

An example absorption profile that results when $\Delta_3$ is scanned is displayed in figure \ref{fig:EIT}b). For $\Delta_3= \Delta_1$, a coherent superposition of state $\ket{1}$ and state $\ket{3}$ is created again as was discussed above, a dark state that does not absorb light.

We have seen that by employing two fields the absorption spectrum can be altered drastically. Such a modified absorption spectrum can be exploited for cooling of ions: The absorption profile is positioned relative to the the sideband spectrum of the ion(s) such that a high absorption probability is obtained for the red sideband (leading to cooling) while the absorption probability for the carrier and the blue sideband (contributing to heating) remains small (see Fig \ref{fig:EIT}b)). With a Ca$^+$ ion, EIT cooling has been demonstrated  \cite{Roos2000}. For a collection of ions that form a Coulomb crystal it should be possible to make the absorption spectrum overlap with the sideband spectrum such that several vibrational modes are cooled simultaneously.

Already this relatively simple system (three atomic states and two fields) allows for modifying the atomic response considerably to achieve efficient laser cooling. Of course, in order to make this qualitative description of EIT cooling quantitative,  terms that describe the atomic motion and spontaneous emission have to be added to the Hamiltonian. A detailed theoretical analysis of EIT cooling of a trapped atom for the case $\Delta_1=\Delta_3$ (two-photon resonance) can be found in \cite{Morigi2003}. There, the cooling rate and and the mean phonon number in steady state are given for an atom confined in a harmonic oscillator  potential.

\section{Cavity cooling}
\index{cavity cooling}
The atomic absorption, emission, and coherent scattering characteristics can be modified strongly by placing an ion in a cavity \cite{Meyer1997,Guthohrlein2001,Mundt2002,Herskind2009} which could be exploited for cooling \cite{Mossberg1991,Cirac1995,Horak1997,Vuletic2000,Beige2005,Zippilli2005}. For example, enhancing atomic scattering into a cavity mode that is tuned to the blue sideband resonance of a trapped ion will lead to cooling of the ion when the incident photon had a lower frequency than the scattered photon \cite{Leibrandt2009}.

\section{Cooling scheme combining laser light and RF}
\index{cooling using laser and rf radiation}
As outlined above, EIT cooling relies on shaping the atomic absorption and emission spectrum using  laser fields that create light-induced states and shift them suitably. In \cite{Retzker2007} a different way of using the ac Stark shift is proposed that also eliminates the carrier transition that otherwise would slow down cooling and increase the steady state temperature. One of the two fields employed there could be an RF field, while for repumping a laser field is used.

When laser radiation is applied to ions that are exposed to a magnetic field gradient, then the effective LDP, eq. \ref{eq:eta_eff}, has two components, one being $\eta$ which is due to the momentum transferred to the ion,  the other being $\kappa$, which stems from the state selective displacement (Fig. \ref{fig:LDP_QM}b). This could be exploited for cooling: If, in addition to an RF field driving an ionic resonance, a laser field is used that induces a Raman transition, and the phase between these fields  is adjusted properly, then
(in a classical picture) the momentum kicks experienced by the ion due to light scattering and the displacement could constructively add up to (de-)excite the ion's motion. Here, only the relative phase between the fields is relevant. The proper quantum mechanical treatment shows that again the carrier and blue sideband transition could be eliminated by interference. A cooling scheme based on this that promises high cooling rates is proposed in \cite{Albrecht2011}.

A discussion of further interesting cooling techniques that have been or could be applied to  trapped atomic and molecular ions (e.g., \cite{Birkl1994,Eschner1995a,Beige2011}) is beyond the scope of this introduction. Let us mention a novel fast cooling scheme that relies on pulsed excitation of ionic resonances to exert properly timed  momentum kicks to the ion in order to cool its motion  \cite{Machnes2010}.

\section*{Acknowledgement}
C.W. thanks Timm F. Gloger for making fine graphs. 

\bibliographystyle{ws-rv-van}

\end{document}